\documentclass[fleqn,12pt]{article}
\usepackage[utf8]{inputenc}
\usepackage[T1]{fontenc}

\usepackage{color}
\usepackage{amsmath,amssymb}
\usepackage{graphicx} 
\usepackage{bm}
\usepackage{here}
\RequirePackage[utf8]{inputenc}
\RequirePackage[english]{babel}

\RequirePackage{mathptmx}   

\usepackage{authblk}
\RequirePackage[left=2cm,%
                right=2cm,%
                top=2.25cm,%
                bottom=2.25cm,%
                headheight=12pt,%
                letterpaper]{geometry}%

\usepackage[authoryear]{natbib}
\usepackage[breaklinks=true]{hyperref}
\usepackage{url}
\usepackage{breakcites}
\usepackage{ifpdf}
\usepackage{times}
\usepackage{sourcesanspro}
\usepackage{newtxmath}
\usepackage{textcomp}%
\usepackage{xcolor}%
\usepackage{hyperref,bm}
\usepackage{subfig}

\definecolor{green}{rgb}{0.4660, 0.6740, 0.1880}

\title{\vspace{-1.5cm}Robust training approach of neural networks for fluid flow state estimations}

\author[1]{Taichi Nakamura}
\author[1,*]{Koji Fukagata}

\affil[1]{Department of Mechanical Engineering, Keio University, Yokohama 223-8522, Japan}
\affil[*]{fukagata@mech.keio.ac.jp}

\date{}
\begin{document}
\maketitle

\begin{abstract}

State estimation from limited sensor measurements is ubiquitously found as a common challenge in a broad range of fields including mechanics, astronomy, and geophysics.
Fluid mechanics is no exception --- state estimation of fluid flows is particularly important for flow control and processing of experimental data.
However, strong nonlinearities and spatio-temporal high degrees of freedom of fluid flows cause difficulties in reasonable estimations.
To handle these issues, neural networks (NNs) have recently been applied to the fluid flow estimation instead of conventional linear methods.
The present study focuses on the capability of NNs to various fluid flow estimation problems
from a practical viewpoint regarding robust training.
Three types of unsteady laminar and turbulent flows are considered for the present demonstration: 1. square cylinder wake, 2. turbulent channel flow, and 3. laminar to turbulent transitional boundary layer.
We utilize a convolutional neural network (CNN) to estimate velocity fields from sectional sensor measurements.
To assess the practicability of the CNN models, physical quantities required for the input and robustness against lack of sensors are investigated.
We also examine the effectiveness of several considerable approaches for model training to gain more robustness against the lack of sensors.
The knowledge acquired through the present study in terms of effective training approaches can be transferred towards practical machine learning in fluid flow modeling.
\end{abstract}

\section{Introduction}
\label{sec1}

State estimation from limited available measurements is a challenging task over a wide range of fields such as engineering, economics, ecology, and biology~\citep{simon2006optimal}.
Among classical methods for state estimation, Bayesian estimation~\citep{bayes1763lii} and least-squares method~\citep{gauss1857theory} can be regarded as fundamental estimation techniques and were used in mathematical astronomy.
Although various methods had been suggested and applied to canonical problems, what makes state estimation famous in terms of practicability was the Apollo program.
The Kalman filter~\citep{K1960} was applied to Apollo's navigation system and successfully took it to the moon~\citep{bar2004estimation, S1970}.
Then, Kalman filter had become popular and various extensions have thus far been developed to deal with more complex applications~\citep{HS1984,WV2000,E2003}.

Including such extensions of Kalman filter, various ideas have recently been utilized in solving problems found in a wide range of fields e.g., disease detection in medical field~\citep{MP2009}, recognition of the external world in robotics~\citep{C2011, MOK2011}, power electronic control and distribution in electrical engineering~\citep{DLEBVDB2006}. 
Among these applications, fluid state estimation has been recognized as a particularly difficult example due to strong nonlinearities and a gigantic number of freedom in space and time.
The state estimation of fluid flows can be applied, among others, to flow control~\citep{B2001,BN2015} and weather forecasting~\citep{WH2007,CB2011}.
To our best knowledge, the first attempt to estimate flow fields was performed by~\cite{AM1988}.
They estimated large-scale structures in turbulent shear flow using linear stochastic estimation (LSE).
However, the LSE has a limitation in estimating finer-scale structures due to its linear operations.

To overcome this limitation, Bewley's group examined the possibilities of four-dimensional variational method or Kalman filters to estimate the turbulent channel flow from wall information~\citep{BMT2001,CHBH2006,CCB2011}.
The Kalman filter estimated the velocity field near the wall relatively well, though the accuracy gets worse far from the wall.
They also reported that the nonlinear Kalman filter outperforms the linear Kalman filter and the ensemble Kalman filter is only able to estimate the entire field.
However, this favorable result with the ensemble Kalman filter can be considered limited according to the additional verification performed by~\cite{SH2017}, which also exhibits the difficulties to estimate the 
entire field of wall-bounded turbulence with high accuracy.

A large number of freedom is also a major hindrance in handling several methods for fluid flow estimation.
Due to the large number of discretization points to represent the fluid flow data, models have to deal with an extremely high-dimensional space from low-dimensional sensor information.
This hindrance can be mitigated by means of some low-dimensional forms of a high-dimensional estimation target~\citep{ES1995,CRT2006,D2006}.
For the fluid flow estimation, proper orthogonal decomposition (POD)-based methods such as Gappy POD~\citep{ES1995} have been considered~\citep{BDW2004,W2006,MBKB2018,MFRFT2020}; however, there is also a limitation of feature extraction from fluid flows because of the linear nature of the method.

As a new method to deal with strong nonlinearity and high degrees of freedom in fluid flows, neural networks (NNs) have recently shown a great potential in different flow problems~\citep{BEF2019,BHT2020,BNK2020,FFT2020,fukami2020sparse,D2021}.
For instance, \cite{GDI2019}~estimated large-scale motions in wall-bounded turbulence as a combination of POD modes and by utilizing a convolutional neural network (CNN)~\citep{LBBH1998}.
They reported that the CNN model successfully estimates large-scale motions and outperforms the POD-based estimation method.
\cite{GESAV2020}~also utilized a CNN to estimate velocity fields of a turbulent channel flow and demonstrated the superiority of CNN over LSE in terms of estimation accuracy.
The comparison between CNN and LSE has also been demonstrated from the viewpoint of noise robustness~\citep{nakamura2021comparison}.
More recently, a generative adversarial network (GAN) has also been considered for state estimations~\citep{GTDISAV2021,KKWL2021}.

\begin{figure}
\begin{center}
\includegraphics[width=0.99\textwidth]{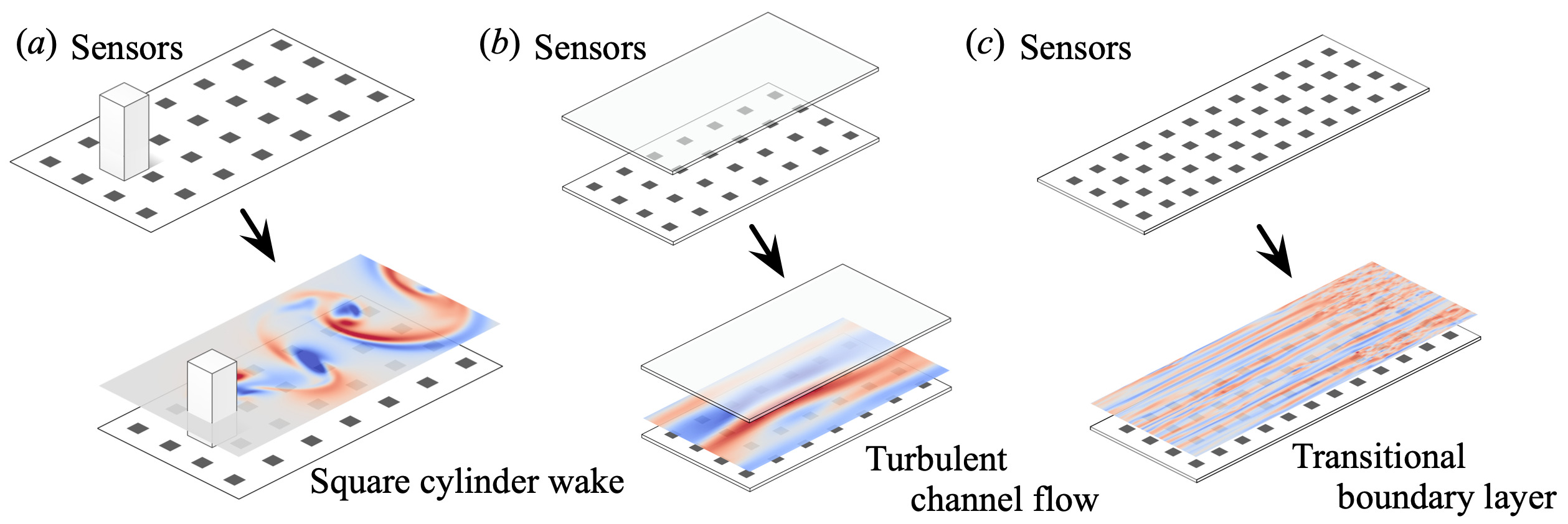}
\caption{
Overview of the velocity estimation problems covered in the present study. $(a)$ Square cylinder wake, $(b)$ Turbulent channel flow, and $(c)$ Transitional boundary layer.
}
\label{fig_overview}
\end{center}
\end{figure}

Since NN has a great potential for fluid flow estimation as introduced above, of particular interest here as the next step is the capability of the NNs from a practical viewpoint.
To this end, this paper discusses the practicability of the NN-based estimation method for fluid flows from the view of robust model construction.
Considering three types of unsteady flows summarized in figure~\ref{fig_overview}, we investigate the dependence of the model performance on physical quantities required for the estimation and the minimum number of sensors.
We also evaluate several approaches for the model to obtain robustness against the lack of sensors.
The present paper is organized as follows.
Details of the estimation model and training data are provided in Sections~\ref{sec2} and~\ref{sec3}.
The estimation performance including the dependence on input attributes and robustness against the lacked input is demonstrated in Section~\ref{sec4-1}.
Several methods to acquire robustness for the lacked input are investigated in Section~\ref{sec4-2}.
We lastly present a summary and provide some outlooks in Section~\ref{sec5}.

\section{Convolutional neural network-based state estimator for fluid flows}
\label{sec2}

As presented in figure~\ref{fig_overview}, we consider state estimation of fluid flows from measurements available distant from the target region.
In this study, both input and output are two-dimensional sections such that the present model attempts to estimate sectional data from sectional inputs.
We perform this sectional estimation capitalizing on a convolutional neural network~\citep{LBBH1998}.
The CNN has recently been identified as one of the promising tools for data-driven fluid flow analyses including state estimation~\citep{FFT2019a,FFT2020b,KSMIM2021}, flow control~\citep{LKBG1997,RKJRC2019}, reduced-order modeling~\citep{MFF2019,HFMF2020a,HFMF2020b,FNF2020,KL2020,nakamura2020extension,MLB2021}, and turbulence modeling~\citep{FNKF2019,DIX2019,LMCVP2019,PSRV2020,TWPH2020,FWNT2021}, thanks to a filter operation inside the CNN~\citep{FFT2020}.

\begin{table}
    \caption{{\color{black}Structure of 2D CNN used in this study. The number of input attributes for the turbulent channel flow or the transitional boundary layer is represented as $N_P\,(=1,2,3)$. ``Conv.'' stands for a convolutional layer.}}
        \centering
        \begin{tabular}{cccccc}\hline
        \multicolumn{2}{c}{Square cylinder} & \multicolumn{2}{c}{Turbulent channel flow} & \multicolumn{2}{c}{Transitional boundary layer} \\
        Layer & Data size & Layer & Data size & Layer & Data size \\
        \hline
        Input (${\bm u}$) & $(256,128,3)$ & Input (${\tau_x}$, ${\tau_z}$, ${p}$) & $(32,32,N_P)$ & Input (${\tau_x}$, ${\tau_z}$, ${p}$) & $(128,128,N_P)$ \\
        1st Conv. & $(256,128,64)$ & 1st Conv. & $(32,32,32)$ & 1st Conv. & $(128,128,32)$\\
        2nd Conv. & $(256,128,64)$ & 2nd Conv. & $(32,32,32)$ & 2nd Conv. & $(128,128,32)$\\
        3rd Conv. & $(256,128,32)$ & 3rd Conv. & $(32,32,32)$ & 3rd Conv. & $(128,128,32)$\\
        4th Conv. & $(256,128,32)$ & 4th Conv. & $(32,32,32)$ & 4th Conv. & $(128,128,32)$\\
        5th Conv. & $(256,128,32)$ & 5th Conv. & $(32,32,32)$ & 5th Conv. & $(128,128,32)$\\
        6th Conv. & $(256,128,32)$ & 6th Conv. & $(32,32,32)$ & 6th Conv. & $(128,128,32)$\\
        7th Conv. & $(256,128,32)$ & 7th Conv. & $(32,32,32)$ & 7th Conv. & $(128,128,32)$\\
        8th Conv. & $(256,128,32)$ & 8th Conv. & $(32,32,32)$ & 8th Conv. & $(128,128,32)$\\
        9th Conv. & $(256,128,32)$ & 9th Conv. & $(32,32,32)$ & 9th Conv. & $(128,128,32)$\\
        10th Conv. & $(256,128,32)$ & 10th Conv. & $(32,32,32)$ & 10th Conv. & $(128,128,32)$\\
        11th Conv. & $(256,128,32)$ & Output (${\bm u}$) & $(32,32,3)$ & 11th Conv. & $(128,128,32)$\\
        12th Conv. & $(256,128,3)$ & & & 12th Conv. & $(128,128,3)$\\
        Output (${\bm u}$) & $(256,128,3)$ & & & Output (${\bm u}$) & $(128,128,3)$\\
        \hline
        &&&&&
        \end{tabular}
        \label{tab1}
\end{table}
The present CNN is composed of convolutional layers, which allows us to extract spatial coherent features of data through filter operations.
Note that pooling or upsampling operations are not considered in the present study because no dimension reduction or expansion is required for the present model such that $\mathbb{R}^{d_{\rm input}}=\mathbb{R}^{d_{\rm output}}$, where $d_{\rm input}$ and $d_{\rm output}$ denote the vector dimensions of input and output data, respectively~\citep{MFZNF2021}.
The operation at the $s$-th convolutional layer $q^{(s)}$ can mathematically be expressed as 
\begin{equation}
    q^{(s)}_{ijn}=\varphi\left(\sum_{m=1}^M\sum_{p=0}^{H-1}\sum_{q=0}^{H-1}h^{(s)}_{pqmn}q^{(s-1)}_{i+p-G,j+q-G,m}+b_n^{(s)}\right),
    \label{eq:CNN}
\end{equation}
where $G=\lfloor H/2\rfloor$, $H$ is width and height of the filter, $M$ is the number of input channels, $n$ is the number of output channels, $b$ is a bias, and $\varphi$ denotes an activation function, respectively.
{\color{black}We set the filter size as $H=3$ in this study.}
Although we can choose a nonlinear activation function from various candidates~\citep{FHNMF2020,MFF2019}, we use ReLU~\citep{NH2010} to avoid vanishing the gradient of weights.
The training of CNN can be regarded as an optimization process regarding weights ${\bm w}$.
The weights are optimized through the backpropagation~\citep{H1992} minimizing a cost function between estimated data and reference data ${\bm q}_{\rm Ref}$,
\begin{equation}
    {\bm w}={\rm argmin}_{\bm w}||{\bm q}^{(s_{\rm max})}-{\bm q}_{\rm Ref}||_2,
    \label{eq2}
\end{equation}
where ${\bm q}^{(s_{\rm max})}$ is the output of CNN at the last layer $s_{\rm max}$.
We use the $L_2$ error norm as the cost function.
The Adam optimizer~\citep{Kingma2014} is utilized to perform the present optimization.

{\color{black}
The structure of the CNN models are summarized in table \ref{tab1}.
Note that we do not consider other model structures because we focus on the robust training approach in this study.
For more information about the effect of model structure on robustness, please refer to \citet{nakamura2021comparison}.
}

\begin{figure}
\begin{center}
\includegraphics[width=0.99\textwidth]{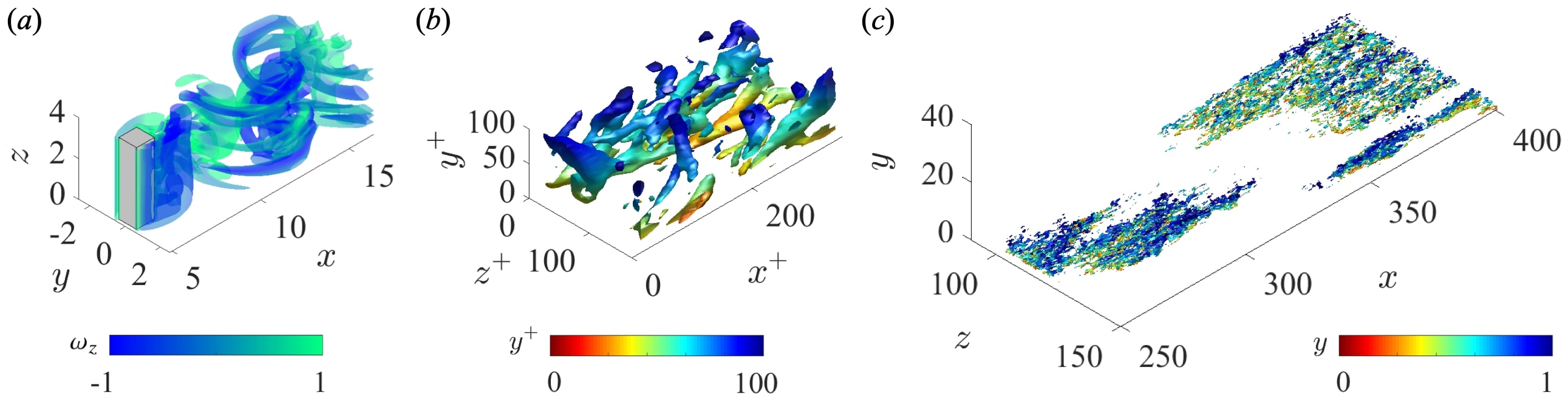}
\caption{
Vortical structures
of the flow fields covered in the present study. $(a)$ Square cylinder wake visualized by the $\lambda_2$ criterion with $\lambda_2=-0.001$, $(b)$ Turbulent channel flow visualized by the $Q$ criterion with $Q^+=0.01$, and $(c)$ Transitional boundary layer ($Q=0.04$).
}
\label{fig_floweg}
\end{center}
\end{figure}
\section{Fluid flow data sets}
\label{sec3}

This study demonstrates the capability of the CNN for fluid flow estimation tasks by considering three types of fluid flow data sets that cover a broad range of spatial length scales of laminar and turbulent flows.
We hereafter introduce the data sets for the present demonstration.

\subsection{Wake around a square cylinder}

A flow around a square cylinder at the Reynolds number ${Re}_D=300$ is first considered.
Although this is a laminar example, the flow at the present Reynolds number can be regarded as a good candidate to discuss the reconstructability of the present CNN model because there are complex three-dimensional structures associated with two- and three-dimensional separated shear layers~\citep{BA2018}, as shown in figure~\ref{fig_floweg}$(a)$.
A direct numerical simulation (DNS) is used to prepare the training data by numerically solving the incompressible Navier--Stokes equations with a penalization term~\citep{volumePenal1994,matsuo2021supervised,MFF2020}, i.e.,
\begin{flalign}
        {\bm{\nabla}} \cdot {\bm u}=0,~~~{\partial_t {\bm u}} + {\bm{\nabla}} \cdot \left({\bm{u}}{\bm{u}}\right)=-{\bm{\nabla}} p + {{ Re}^{-1}_D}{\bm{\nabla}}^2\bm{u}+\lambda \chi\left({\bm u}_b-{\bm u}\right),
\end{flalign}
where ${\bm u}=\{u,v,w\}$ and $p$ are the velocity vector and pressure, which are nondimensionalized with the fluid density $\rho$, the length of the square cylinder $D$, and the uniform velocity $U_\infty$.
The penalization term expresses the bluff body with a penalty parameter $\lambda$, a mask value $\chi$, and the velocity of the object ${\bm u}_b$ which is zero in the present case.
For the mask value, $\chi=0$ and 1, respectively corresponding to outside and inside a body.
The spatial domain in the present simulation covers $\left(L_x, L_y, L_z\right)=\left(20D,20D,4D\right)$. 

The DNS code is based on that developed for turbulent channel flows by \citet{FKK2006}: the spatial discretization is done by using the energy-conservative second-order finite difference method, and the time integration is done by using the low-storage third order Runge-Kutta/Crank-Nicolson method.
The computation is carried out with the time step of $\Delta t=2.5\times10^{-3}$.
A uniform velocity is imposed at the inflow boundary, while the convective boundary condition is applied at the outflow boundary.
We consider the slip boundary condition at $y=0$ and $y=L_y$, and the periodic boundary condition at $z=0$ and $z=L_z$.
The center of the square cylinder is located $5.5D$ downstream of the inflow boundary.

For the present training data, the part of computational volume around the square cylinder $\left(12.8D,4D,4D\right)$ with the grid number of $(N_x^\sharp, N_y^\sharp, N_z^\sharp)=(256,128,160)$ is extracted, and 1000 snapshots are prepared.
We choose 70\% of the snapshots for the training, while remaining 30\% is used for the validation.
We also consider additional 1000 snapshots for the assessment in Section~\ref{sec4-1-1}.
For this square cylinder example, we use the velocity vector ${\bm u}=\{u,v,w\}$ as the qualities of interest for both input and output.

\subsection{Turbulent channel flow}

Similar to out previous work on CNN autoencoder \citep{nakamura2020extension},
a minimal turbulent channel flow~\citep{JM1991} at $Re_{\tau}=110$ is then considered, as shown in figure \ref{fig_floweg}$(b)$.
The training data are obtained by a DNS which numerically solves the incompressible continuity and Navier--Stokes equations,
\begin{equation}
    \bm{\nabla} \cdot {\bm u} = 0,~~~{ {\partial_t {\bm u}} + \bm{\nabla} \cdot ({\bm u \bm u}) =  -\bm{\nabla} p  + {{Re}^{-1}_\tau}\nabla^2 {\bm u}},
\end{equation}
where ${\bm u}$ and $p$ represents the velocity vector and pressure, respectively.
The quantities used in the equations are nondimensionalized with the channel half-width $\delta$ and the friction velocity $u_\tau$.

The computational domain covers $(L_{x}, L_{y}, L_{z}) = (\pi\delta, 2\delta, 0.5\pi\delta)$ with the grid numbers of $(N_{x}, N_{y}, N_{z}) = (32, 64, 32)$.
A uniform grid is used in the streamwise ($x$) and the spanwise ($z$) directions, while a nonuniform grid being used in the wall-normal ($y$) direction.
The numerical scheme for DNS is exactly the same as that for the square cylinder case.
The time step is set to $\Delta t^+=0.0385$, where the subscript $+$ denotes the wall units.

We use 10000 snapshots for the training of the present CNN.
We use 70\% of the snapshots for the training, and the remaining 30\% is used for the validation.
Note that the channel flow is considered for both the comparison among various fluid flow data sets (section~\ref{sec4-1}) and the investigation with regard to acquisition of noise robustness in a training framework (section~\ref{sec4-2}). 
For both assessments, additional 5000 snapshots are prepared.
Although details will be provided later, the present CNN for the turbulent channel flow example attempts to estimate the velocity vector ${\bm u}=\{u,v,w\}$ from wall measurements.
The dependence of the estimation accuracy on the choice of input quantities will also be investigated.

\subsection{Transitional boundary layer}

As a more complex example, we also consider a transitional boundary layer prepared from Johns Hopkins Turbulence Databases~\citep{LPWYMBCSE2008, PBLM2007}.
For details of the computational conditions, please refer to~\cite{Z2013}.
The data sets are obtained by DNS of incompressible flow over a flat plate with an elliptical leading edge. 
The Reynolds number based on the plate half-thickness $L$ is $Re_{L}=800$.
The computational domain non-dimensionalized by $L$ is $(L_x,L_y, L_z)=(1099, 40, 240)$ and the simulation time step is $\Delta t = 0.005$.
A no-slip boundary condition is applied at the wall.

From the original computational domain of the database, we extract the transition region defined by the momentum-thickness Reynolds number so that the extracted domain contains both laminar and turbulent structures, as shown in figure~\ref{fig_floweg}$(c)$.
The extracted domain size is $(149.7, 26.4, 60.0)$ with the number of grid points of $(N_x,N_y,N_z)=(128,224,128)$.
Based on the above configuration, we prepare 700 snapshots, 70\% of which is for the training and the remaining 30\% is for the validation.
We also prepare additional 250 snapshots as the test data for the assessment in Section~\ref{sec4-1-3}.
Similar to the channel flow case, the present CNN with this transitional example also aims to estimate the velocity field ${\bm u}=\{u,v,w\}$ from wall measurements.

\section{Results}
\subsection{Demonstration of CNN-based state estimator for unsteady laminar and turbulent flows}
\label{sec4-1}
\subsubsection{Square cylinder wake}
\label{sec4-1-1}

\begin{figure}[t]
\begin{center}
\includegraphics[width=0.99\textwidth]{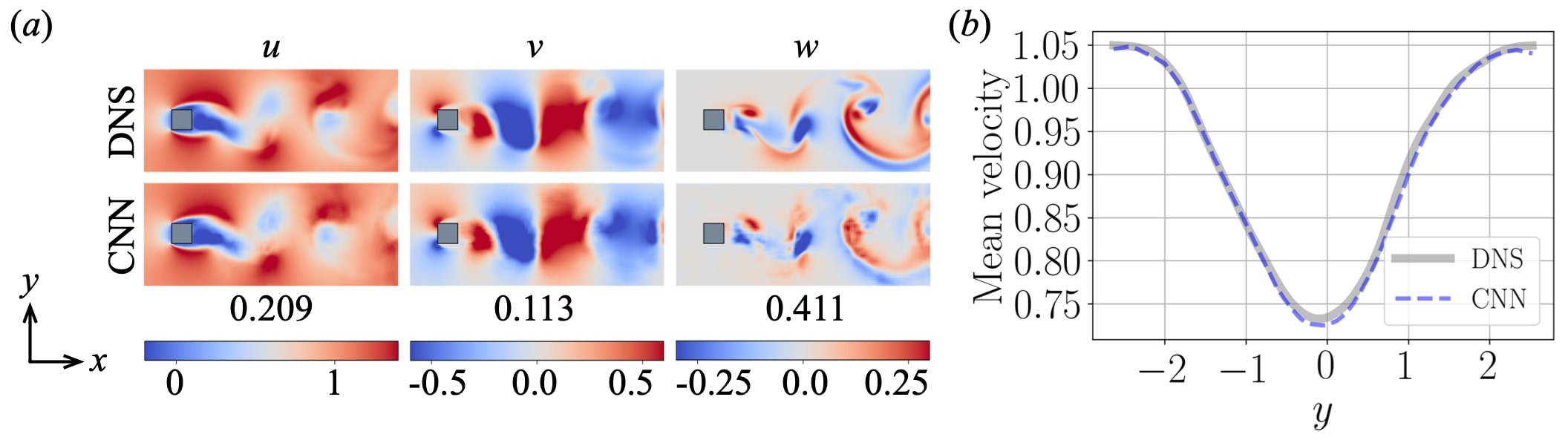}
\caption{
Estimation of a square cylinder wake. $(a)$ Velocity fields and the $L_2$ error norms. $(b)$ Mean streamwise velocity profile at $x=10.0$.
}
\label{fig_base-sqcy}
\end{center}
\end{figure}


\begin{figure}[t]
\begin{center}
\includegraphics[width=0.98\textwidth]{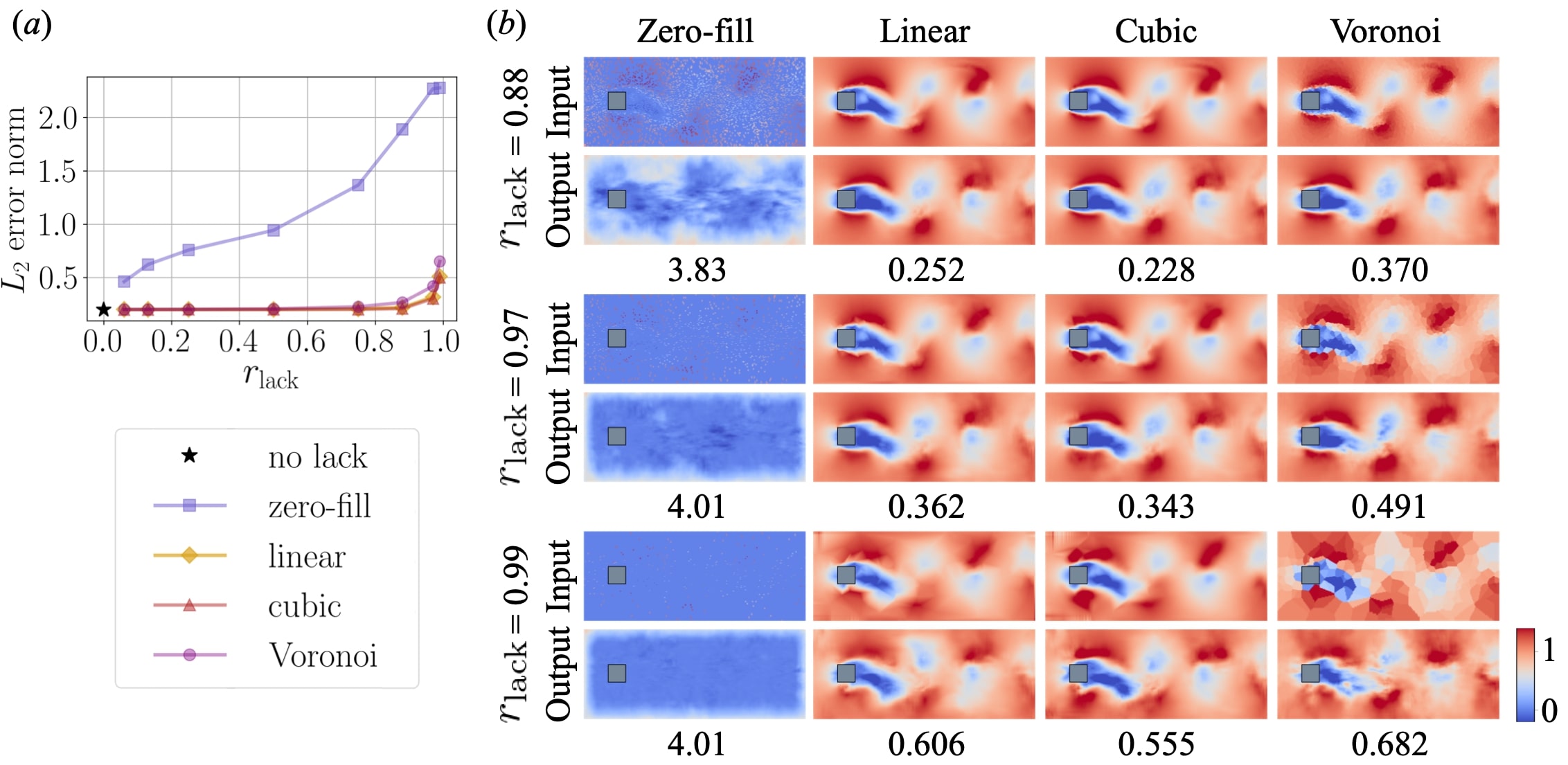}
\caption{{\color{black}
Robustness of the CNN model against the lack of sensors for the estimation of a square cylinder wake. $(a)$ Relationship between the $L_2$ error norm and the lack ratio. $(b)$ Streamwise velocity fields of input and output data. The values beneath the contours represent the $L_2$ error norm.}}
\label{fig_lack_sqcy}
\end{center}
\end{figure}

Let us first apply the CNN to a square cylinder wake at $Re_D=300$.
The present CNN model ${\mathcal F}$ estimates an $x-y$ sectional velocity field ${\bm u}$ at $z=1.5$ (non-dimensionalized by $D$) from velocity sensors ${\bm s}$ collected from the $x-y$ cross-section at $z=2.0$ such that ${\bm u}_{z=1.5}={\mathcal F}({\bm s}_{z=2.0})$.
The estimated velocity fields with the $L_2$ error norm are presented in figure~\ref{fig_base-sqcy}$(a)$.
Here, the $L_2$ error norm $\epsilon$ is defined as $\epsilon = {||{\bm u}_{\rm ML} - {\bm u}_{\rm DNS}||_2}/{||{\bm u}^{\prime}_{\rm DNS}||_2}$, where ${\bm u}_{\rm ML}$ is the estimated velocity field, ${\bm u}_{\rm DNS}$ is the reference DNS field, and ${\bm u}^{\prime}_{\rm DNS}$ denotes the velocity fluctuations of DNS data.
The reconstructed velocity fields are in reasonable agreement with the DNS data.
The reasonable reconstruction can also be observed with the streamwise mean velocity profile in figure~\ref{fig_base-sqcy}$(b)$.

We then investigate the capability of the present CNN model from a practical viewpoint.
Although we used information from all grid points on the cross section as the input, which means we have to arrange sensors without gaps --- this is not realistic.
Hence, the robustness of the CNN model against the lacked input data is examined here.
The sensor placements are randomly determined, and this information is fed into the CNN model already trained with the full data.
However, since two-dimensional CNN can only handle sectional data (i.e., not local sensor measurements), an appropriate preprocessing is required to feed the sensor information into the CNN directly.
To do this, we consider four methods to treat the lack of input: 1. zero-fill (i.e., substituting zero to the lacked part), 2. linear interpolation, 3. cubic interpolation, and 4. Voronoi tessellation.
The Voronoi tessellation~\citep{V1908} can handle random sensor placements, and its usefulness for fluid flow data has been demonstrated by~\cite{FukamiVoronoi}.
{\color{black}Note that for the linear and cubic interpolation, the edge region not surrounded by the sensors is filled by linear extrapolation.}

The error for each ratio of lacked data $r_{\rm lack}=n_{\rm lack}/n_{\rm all}$ is presented in figure~\ref{fig_lack_sqcy}$(a)$.
The number of all sensors in this problem corresponds to the number of grid points in DNS, i.e., $n_{\rm all}=256\times128=32768$, while $(n_{\rm all}-n_{\rm lack})$ is the number of randomly placed sensors used for the input.
As shown, the error of the interpolation methods is quite smaller than that of zero-fill, which implies the effectiveness of interpolation to keep the robustness.
The estimated velocity fields from the lacked input are also visualized in figure~\ref{fig_lack_sqcy}$(b)$.
{\color{black}
We can observe the superiority of the cubic interpolation method for all of the lack ratios.
The estimated fields using cubic interpolation with the lack ratio $r_{\rm lack}$ of $\{0.88, 0.97\}$ are in qualitative agreement with the reference DNS data, though the wake structures slightly deform or disappear at $r_{\rm lack}=0.99$.
Therefore, the present CNN model for a square cylinder wake can keep the robustness for as much as $r_{\rm lack}=0.97$ when the cubic interpolation is utilized.}

\subsubsection{Turbulent channel flow}
\label{sec4-1-2}

Let us then use a turbulent channel flow as a complex flow example.
The CNN models ${\mathcal F}$ estimate the velocity field $\{u,v,w\}$ at $y^+=15.4$ from sensor measurements on the wall such that ${\bm u}_{y^+=15.4}={\mathcal F}({\bm s}_{\rm wall})$.
From the perspective of saving sensors, it is important to know physical quantities that can contribute to estimations.
Hence, we here examine the dependence of the model performance on the input attributes.
As the candidates for physical quantities, we consider several combinations of three physical quantities, streamwise and spanwise wall shear stress $\tau_x, \tau_z$ and pressure $p$, which are often used for state estimation in turbulent channel flow~\citep{SH2017,GESAV2020}.

\begin{figure}
\begin{center}
\includegraphics[width=1.00\textwidth]{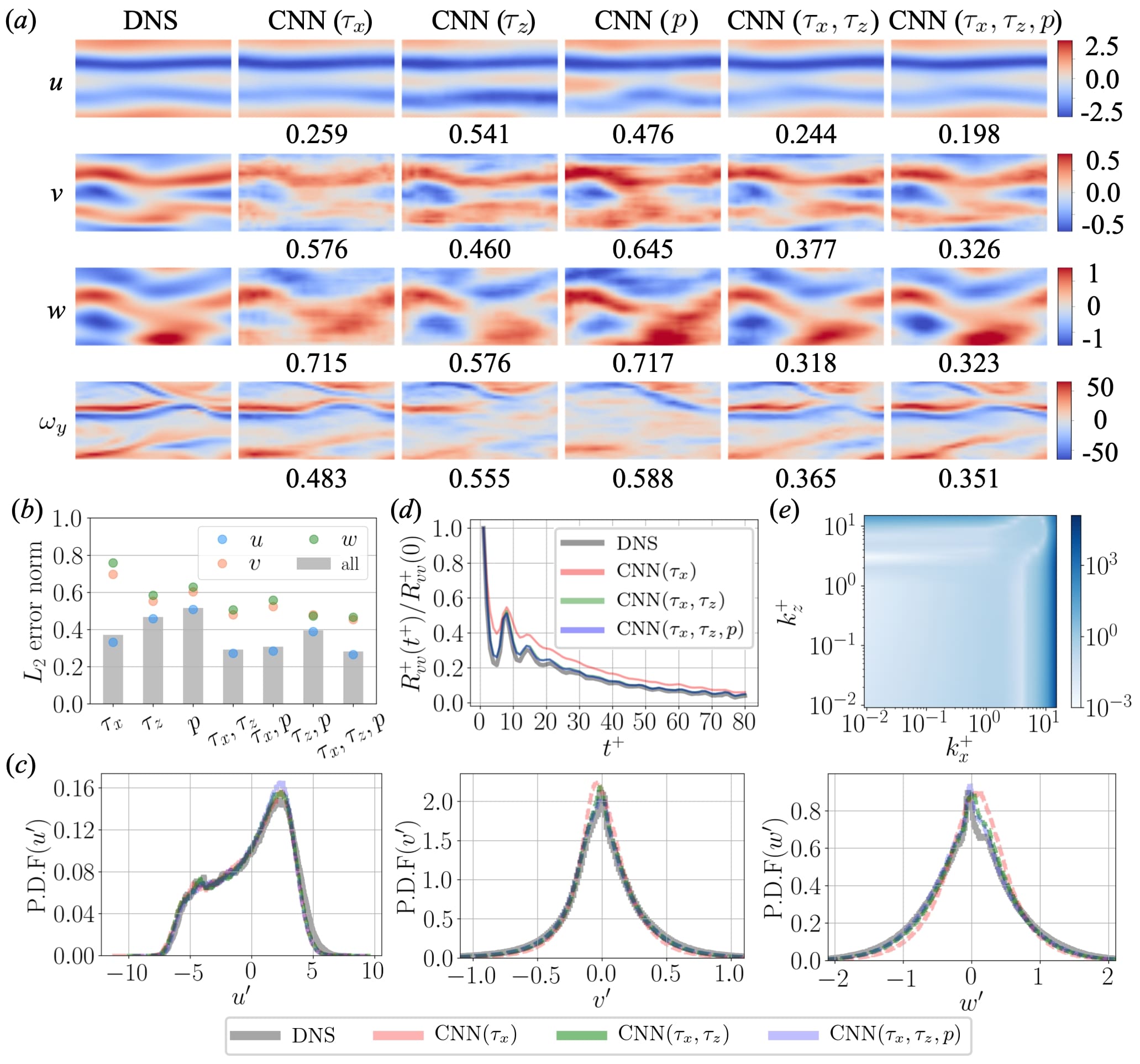}
\caption{
Dependence of the model performance on input attributes for turbulent channel flow. $(a)$ Velocity and vorticity fields. The values beneath the contours represent the $L_2$ error norm. $(b)$ $L_2$ error norm of each model. ``All" represents an ensemble error over three velocity components. $(c)$ Probability density function of each velocity component. $(d)$ Temporal two-point correlation coefficient. $(e)$ $L_2$ error map of the energy spectrum between the DNS and the field reconstructed with the input of $\{\tau_x, \tau_z\}$.}
\label{fig_input-CHNL}
\end{center}
\end{figure}

The estimation performance of each input case is summarized in figure~\ref{fig_input-CHNL}.
Note that the vorticity field $\omega_y$ in figure~\ref{fig_input-CHNL}$(a)$ is obtained from the estimated velocity fields $u$ and $w$ such that
$\omega_{y, {\rm ML}}=f(u_{\rm ML},w_{\rm ML}) = {\partial u_{\rm ML}}/{\partial z} - {\partial w_{\rm ML}}/{\partial x}.$
Hence, the assessment on $\omega_{y,{\rm ML}}$ can be regarded as tougher than the use of velocities only because the first-order differential needs to be calculated from the machine-learned velocities.
The reconstructed velocity and vorticity fields are in reasonable agreement with the reference DNS.
The $L_2$ error is also summarized in figure~\ref{fig_input-CHNL}$(b)$.
Among the cases with a single quantity input i.e., $\{\tau_x\}$, $\{\tau_z\}$, and $\{p\}$, the $L_2$ error with the streamwise wall shear stress $\tau_x$ is the smallest, which suggests that ${\tau_x}$ most significantly contributes to the estimation.
This can particularly be found from the vorticity contours $\omega_y$ in figure~\ref{fig_input-CHNL}$(a)$.
The model with $\{\tau_x\}$ input can estimate the DNS-like structures well, while the models with the input of $\{\tau_z\}$ and $\{p\}$ cannot estimate vortex structures. 
However, the $L_2$ error of $\{\tau_x\}$ input only for $v$ and $w$ is larger than that of $\tau_z$.
This implies that the reasonable estimation for all velocity components requires both $\tau_x$ and $\tau_z$.
The necessity of the spanwise wall shear stress input $\{\tau_z\}$ can also be observed with the probability density function in figure~\ref{fig_input-CHNL}$(c)$.
For $v^{\prime}$ and $w^{\prime}$, the curve of the CNN with $\{\tau_x\}$ input is a little shifted from that of the DNS, while that of the CNN with $\{\tau_x, \tau_z\}$ input being in reasonable agreement with the DNS.
Although we investigate the influence of the pressure input, it contributes to the estimation a bit, and the performance of the model with $\{\tau_x, \tau_z, p\}$ input is not so different from that with $\{\tau_x, \tau_z\}$ input.
We also assess the reconstruction in temporal behavior, considering temporal two-point correlation coefficient $R^+_{vv}(t^+)/R^+_{vv}(t^+=0)$, as presented in figure~\ref{fig_input-CHNL}$(d)$.
The coefficient is defined as
\begin{equation}
    R^+_{vv}(t^+) = {\overline{v^\prime(t_0^++t^+,x,z)v^\prime(t_0^+,x,z)}^{t_0^+,x,z}}.
\end{equation}
The curves for the models $\{\tau_x, \tau_z\}$ and $\{\tau_x, \tau_z, p\}$ are in good agreement with the DNS.
With the other assessments above, the input attribute of $\{\tau_x, \tau_z\}$ is sufficient for estimation from the viewpoint of temporal behavior.

\begin{figure}
\begin{center}
\includegraphics[width=0.98\textwidth]{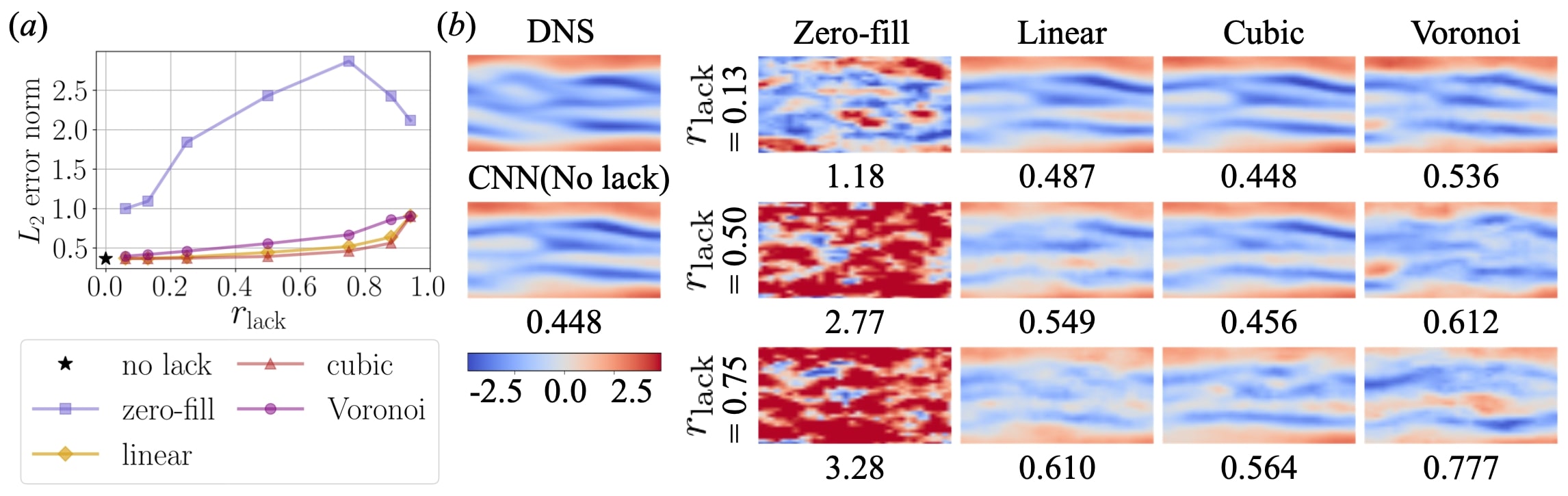}
\caption{{\color{black}
Robustness of the CNN model against the lack of sensors for turbulent channel flow example. $(a)$ Relationship between the $L_2$ error norm and the lack ratio. $(b)$ Dependence of the velocity reconstruction ($u^\prime$) on the interpolation methods and the lack ratio. The values beneath the contours represent the $L_2$ error norm. }}
\label{fig_lack_chnl}
\end{center}
\end{figure}

Let us further assess the capability of the model on the wavespace by focusing on the case with $\{\tau_x, \tau_z\}$ input.
The normalized $L_2$ error map of energy spectrum in the streamwise and spanwise directions is presented in figure~\ref{fig_input-CHNL}$(e)$.
Here, the two-dimensional energy spectrum is defined as
\begin{equation}
    E_{uu}(k_x,k_z) = \overline{\hat{u}^*\hat{u}}^t,
\end{equation}
where $\hat{(\cdot)}$ represents two-dimensional Fourier transformation and ${(\cdot)^*}$ denotes the complex conjugate.
For both the streamwise and spanwise directions, the error is small up to the wavenumber of $10^{0.5}$.
This implies that the CNN model can make physically reasonable estimations although much finer-scale structures cannot be estimated.
This is likely due to the less correlation in the dissipation range~\citep{SSSWPB2020,FHNMF2020}.

\begin{figure}[t]
\begin{center}
\includegraphics[width=0.98\textwidth]{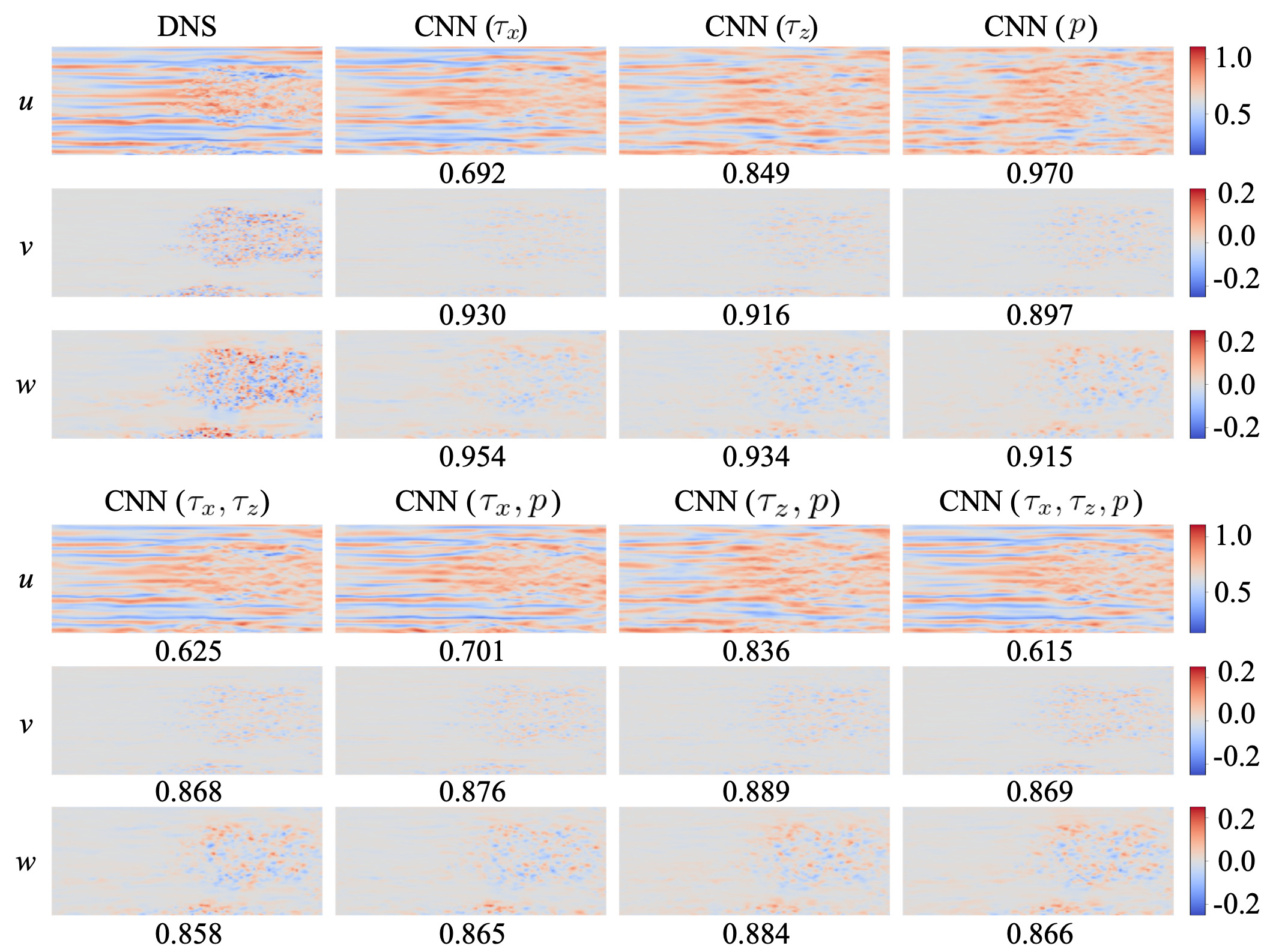}
\caption{
Dependence of the model performance on input attributes for transitional boundary layer. The values underneath the contours represent the $L_2$ error norm.}
\label{fig_input-TBL}
\end{center}
\end{figure}

\begin{figure}[t!]
\begin{center}
\includegraphics[width=1.00\textwidth]{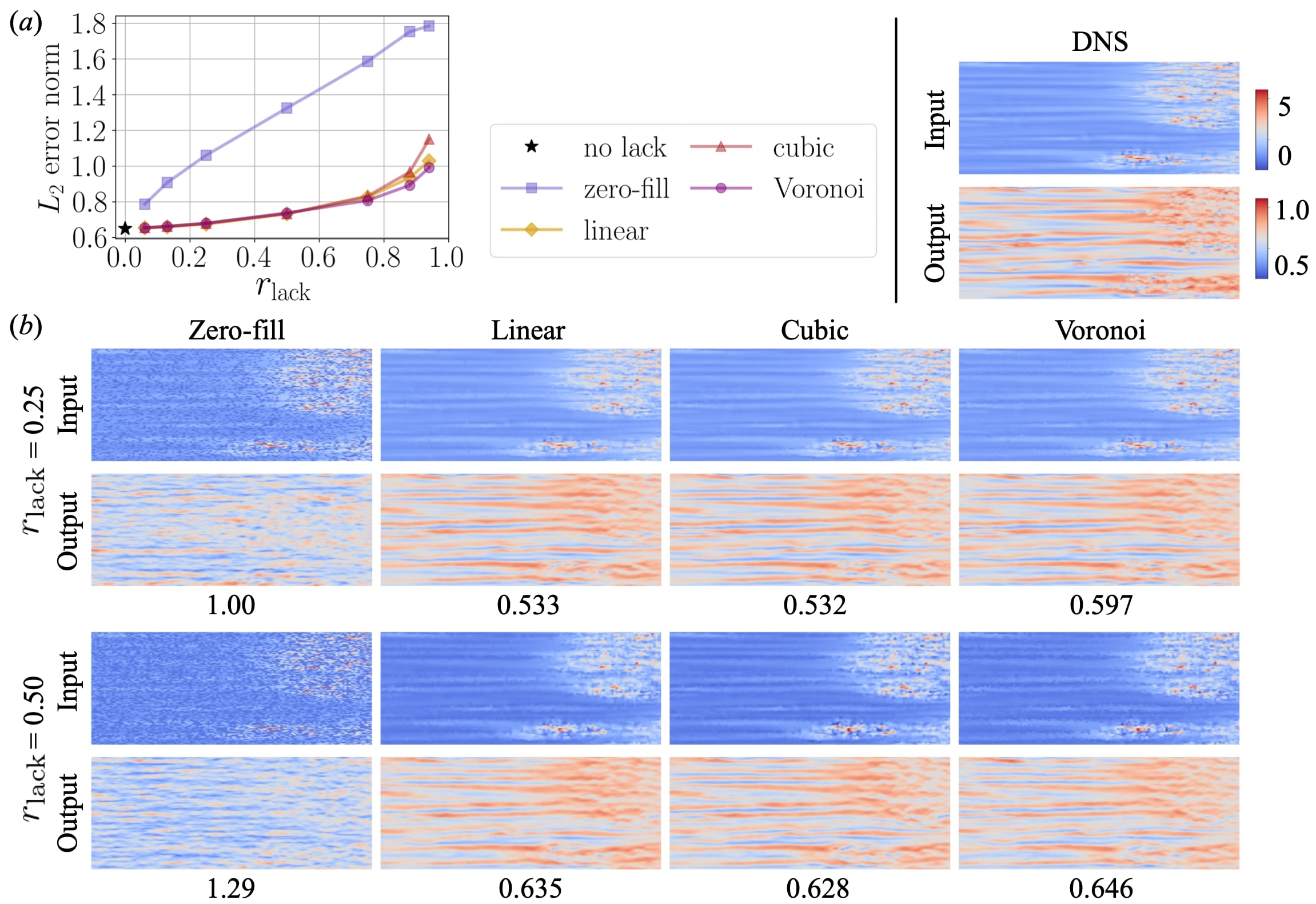}
\caption{{\color{black}
Robustness of the CNN model against the lack of sensors for transitional boundary layer.
$(a)$ Relationship between the $L_2$ error norm and the lack ratio. $(b)$ Dependence of the streamwise velocity reconstruction on the interpolation methods and the lack ratio. The values beneath the contours represent the $L_2$ error norm.}}
\label{fig_lack_tbl}
\end{center}
\end{figure}

\begin{figure}[t!]
\begin{center}
\includegraphics[width=0.95\textwidth]{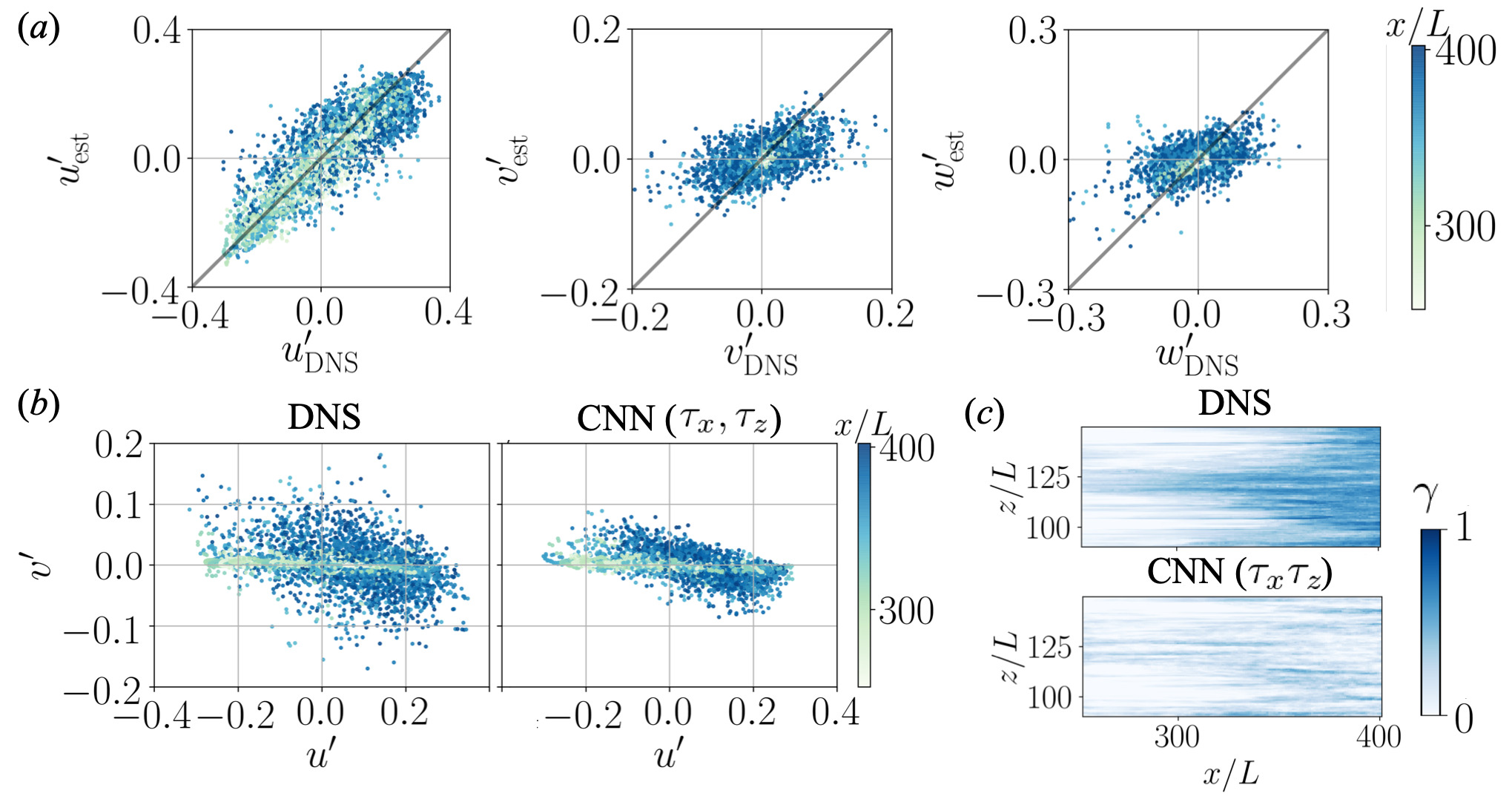}
\caption{
Dependence of the model performance for the example of transitional boundary layer on the streamwise position. $(a)$ 45-degree map of the reference DNS and the estimated field. $(b)$ Joint probability density function of the DNS and the CNN. $(c)$ Intermittency factor map of the DNS and the CNN.
}
\label{fig_TBL-add}
\end{center}
\end{figure}
\begin{figure}
\begin{center}
\includegraphics[width=0.98\textwidth]{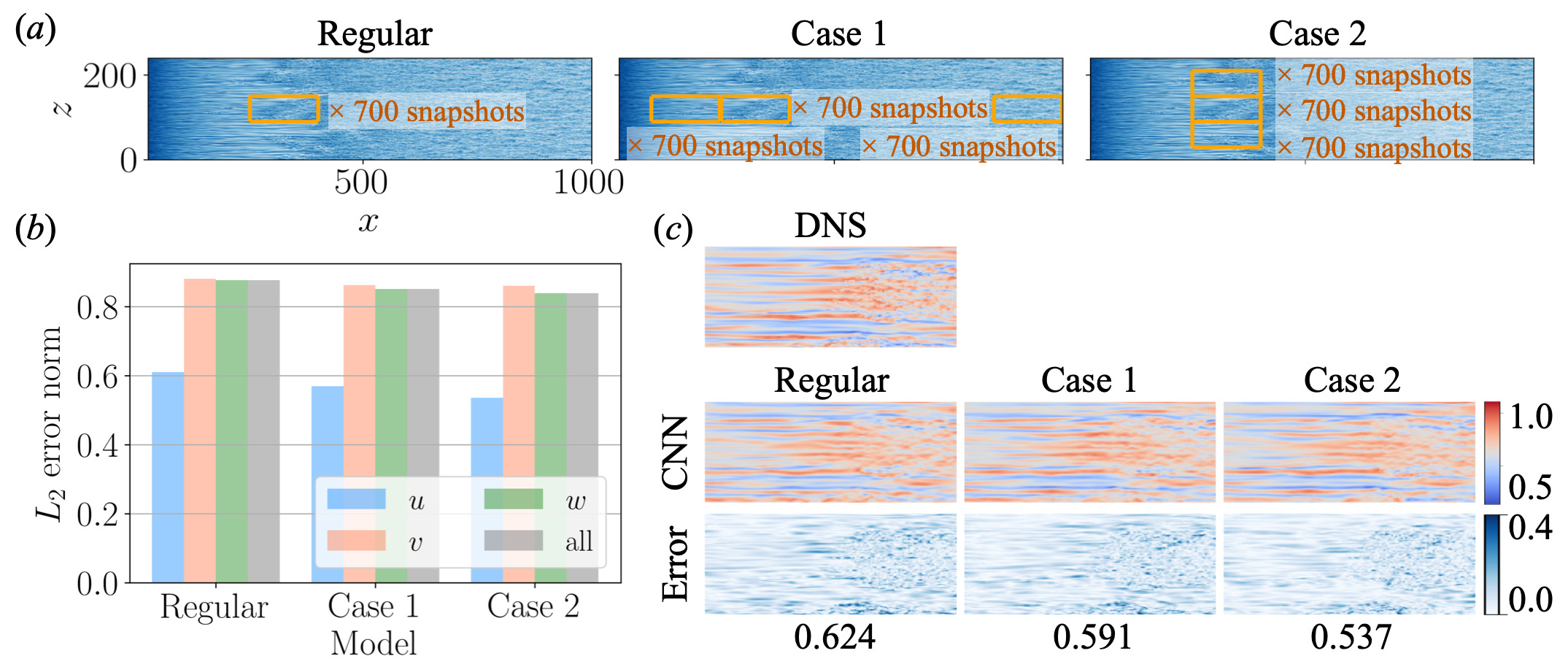}
\caption{Assessment of the CNN models trained with augmented data sets. $(a)$ The regions of each training data. $(b)$ $L_2$ error norm of each model. $(c)$ Velocity contours and $L_1$ norm error map. The values underneath contours represent $L_2$ error norm of the snapshot.}
\label{fig_TBL-imp}
\end{center}
\end{figure}

We also investigate the robustness against the lacked input data of the turbulent channel flow example, analogous to the square cylinder example.
Following the discussion above, we here use the CNN model with $\{\tau_x, \tau_z\}$ input.
The error for each ratio of lack is shown in figure~\ref{fig_lack_chnl}$(a)$.
The smaller error with the interpolation methods than that with zero-fill indicates the effectiveness of interpolation, similarly to the square cylinder example.
{\color{black}We then check the velocity contours to compare the interpolation methods and to see what percentage of lack can be tolerated, as summarized in figure~\ref{fig_lack_chnl}$(b)$.
The superiority of the cubic interpolation can be again observed.
For $13\%$ lack, the estimated flow field is in qualitative agreement with the DNS, whereas the fine structure is lost with $50\%$ lack, and the estimated structure becomes substantially different from that of DNS with $75\%$ lack.
Therefore, we can conclude that the cubic interpolation is the best for preprocessing and the present CNN model for the turbulent channel flow estimation can accept at most $50\%$ lacked input data.}

\subsubsection{A transitional boundary layer}
\label{sec4-1-3}

As a more complex problem, let us apply the present CNN model to the transitional boundary layer flow.
The model $\mathcal{F}$ estimates the velocity field $\{u,v,w\}$ at $y=0.96L$ from sensor information on a flat plate such that ${\bm u}_{0.96L}=\mathcal{F}({\bm s}_{\rm plate})$.
Analogous to the channel flow example, we investigate what input attributes contribute to the transitional boundary layer estimation.
We consider seven cases that are combinations of the streamwise and spanwise shear stresses $\tau_x, \tau_z$ and pressure $p$.
The estimated velocity fields and the $L_2$ error norm are summarized in figure~\ref{fig_input-TBL}.
As clearly seen, the models which include the input of streamwise shear stress $\tau_x$ show the better performance than that without $\tau_x$ input.
However, we should note that the input of $\tau_x$ only is insufficient for the estimation of $v$ and $w$.
Hence, the additional information such as $\tau_z$ and $p$ are required to enhance the estimation accuracy for $v$ and $w$, as presented in figure~\ref{fig_input-TBL}.

We also investigate the robustness against the lacked input, as shown in figure~\ref{fig_lack_tbl}.
{\color{black}The effectiveness for the use of interpolation can be again observed.
We cannot observe a significant difference between the interpolation methods in figure~\ref{fig_lack_tbl}$(a)$, which makes us compare in figure \ref{fig_lack_tbl}$(b)$.
The $L_2$ error norm in figure~\ref{fig_lack_tbl}$(b)$ indicates that cubic interpolation is slightly better than the other methods.
Note that the larger error of cubic interpolation in the range of $r_{\rm lack}\gtrsim 0.75$ in figure~\ref{fig_lack_tbl}$(a)$ does not make sense because the error is quite large.}
The lack acceptability of the models is then examined by observing flow fields in figure~\ref{fig_lack_tbl}$(b)$.
For $25\%$ lack, the overall trends of the estimated field can be kept comparing to the DNS. 
On the other hand, we gradually start to see different flow structures compared to the DNS with $50\%$ lack.
Therefore, the present CNN model for the transitional boundary layer example can accept at most $25\%$ lacked input data.
However, we should note that the error of ``no lack" is originally large, which makes us suspect what the origin of the error is.

As an additional assessment to clarify the point above, the maps of estimated values versus the reference DNS values, usually called 45-degree map, are shown in figure~\ref{fig_TBL-add}$(a)$.
We here consider the CNN model with $\{\tau_x, \tau_z\}$ input.
A large error can be found in the downstream region for the all velocity components.
The same trend can be seen from the comparison of joint probability density function in figure~\ref{fig_TBL-add}$(b)$.
The difference between the upper and downstream region can also be assessed from a physical viewpoint by introducing intermittency factor $\gamma$ defined as the fraction of time where the flow in a given region is turbulent.
Continuous laminar and turbulent flows respectively correspond to $\gamma=0$ and 1.
There are some criteria to determine whether the specific region is turbulent or not.
We here use a modified turbulent energy recognition algorithm (M-TERA) method~\citep{ZCW1995}.
The M-TERA method determines a region of turbulent when it satisfies the following equation:
\begin{equation}
    \overline{\left|u^{\prime}\frac{\partial u^\prime}{\partial t}\right|} > C\left[\overline{u}\frac{(\partial u^\prime/\partial t)_{\rm rms}}{(u^\prime \partial u^\prime /\partial t)_{\rm rms}}\right],
\end{equation}
where $\overline{(\cdot)}$ represents the mean over a short time-interval and $(\cdot)_{\rm rms}$ denotes the long-time standard deviation.
The contours of intermittency factor are presented in figure~\ref{fig_TBL-add}$(c)$.
As can be seen, the downstream region estimated by the CNN model is almost determined as laminar, despite that the region is almost turbulent in DNS.
One of the candidates to improve the estimation performance in the downstream region is training data addition (so-called data augmentation in the field of machine learning~\citep{SK2019,MFZF2020}).
Hence, our next interest is what kind of structures should be contained in the additional training data.

To clarify this point, we here consider two types of training data addition, as presented in \ref{fig_TBL-imp}$(a)$;
\begin{enumerate}
    \item Add 700 snapshots from both laminar and turbulent region to the original 700 snapshots on transient region such that 2100 snapshots in total (Case 1).
    \item Add 700 snapshots from both upper and lower portion on transient region such that 2100 snapshots in total (Case 2). 
\end{enumerate}
We use the streamwise and spanwise wall shear stresses $\tau_x, \tau_z$ as the input attribute for the CNN model.
The performance of each model is then assessed using the test data from the regular region, as summarized in figures~\ref{fig_TBL-imp}$(b)$ and $(c)$.
Both cases show improvements in reconstruction, especially in the $u$ component.
In this particular example, the use of case 2 can affect the accuracy more than that of case 1, which suggests that the transient structures are more important than the laminar/turbulent structures as the additional data in the transient flow estimation.
In this way, we can learn considerable paths to improve the estimation accuracy with preparing proper data augmentation methods.
We also note that the reason why we cannot find the significant improvement for the $v$ and $w$ components is merely because the training data sampling is determined based on the structural distribution of streamwise velocity $u$, as shown in figure~\ref{fig_TBL-imp}$(a)$.
Hence, the use of the $v$ and $w$ components for the training data sampling or incorporating intelligence data preparation~\citep{sapsis2020output} could promote the estimation capability of the model more.

\subsection{Investigations of several methods to acquire robustness for the lacked input with turbulent channel flow}
\label{sec4-2}

In this section, let us introduce the capability of several methods to construct a robust CNN model against lacked input data, especially in fluid flow estimation.
{\color{black}We aim to construct a robust model while maintaining estimation performance for no-lacked data.}
We here consider three methods: 1. regularization, 2. dropout, and 3. noise-perturbed data training.
For the demonstrations in what follows, we use the turbulent channel flow example with the input of streamwise wall shear stress $\tau_x$.

\subsubsection{Regularization}

Regularization is often used in the fields of machine learning and statistics as a method of preventing overfitting and gaining robustness~\citep{SSBO2002}.
This method adds a penalization term to the loss function as
\begin{eqnarray}
    {\bm w} = {\rm argmin}_{\bm w}[||{\bm q}_{\rm ML}-{\bm q}_{\rm Ref}||_A+\alpha||{\bm w}||_B],
\end{eqnarray}
where $\alpha$ is a hyperparameter to determine the magnitude of the regularization, $A$ and $B$ respectively indicate the norm factor for the loss and the regularization terms.
In this study, a set of $\{A, B\}=\{2,[1,2]\}$ is considered, where $B=1$ and $2$ correspond to Lasso ($L_1$ regularization)~\citep{T1996} and Ridge ($L_2$ regularization)~\citep{HK1970}, respectively.

\begin{figure}[t!]
\begin{center}
\includegraphics[width=0.98\textwidth]{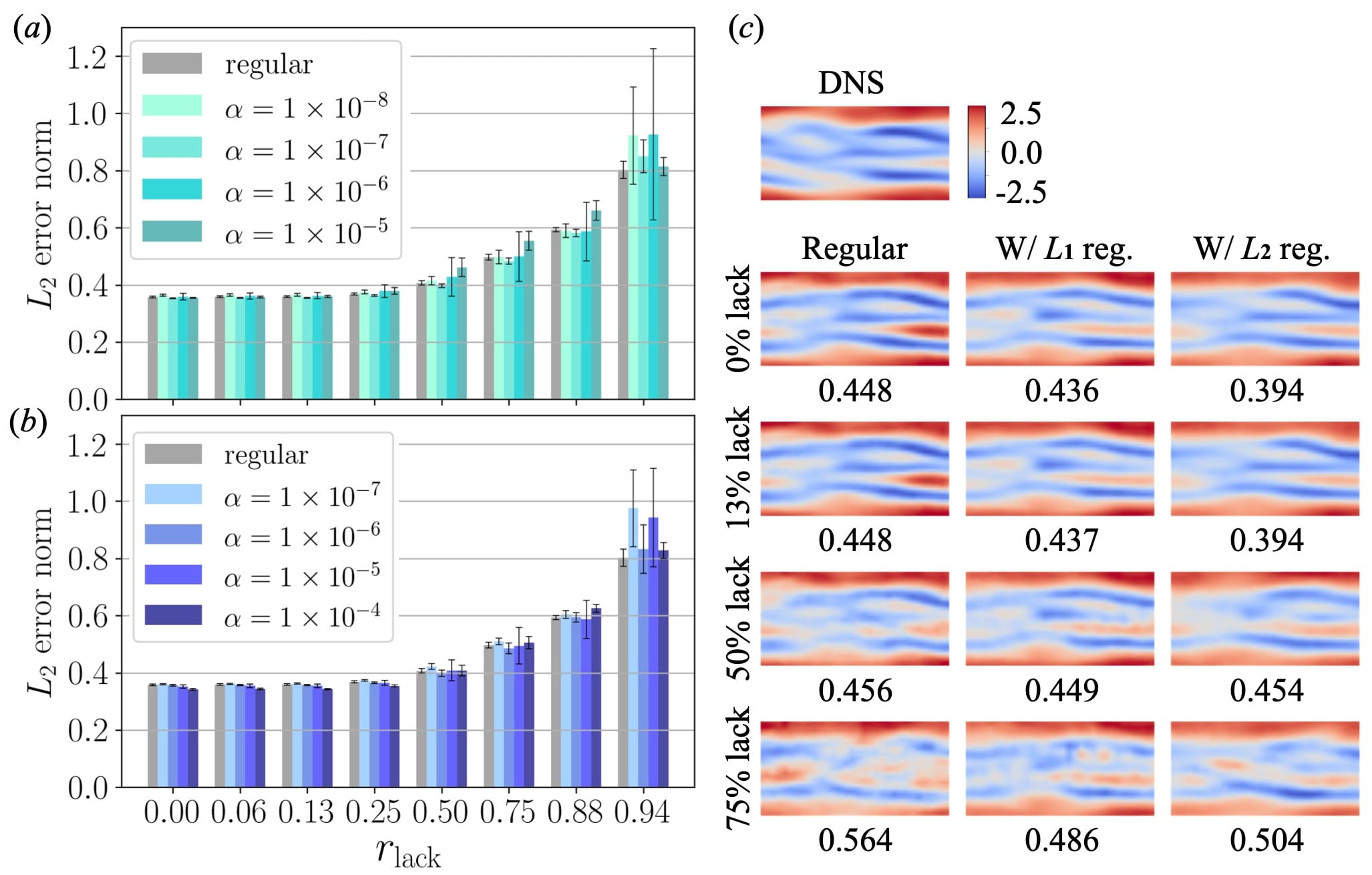}
\caption{{\color{black}
Effect of regularization for the robustness. The ``regular" is the model without regularization. $L_2$ error norm of the model with $(a)$ $L_1$ and $(b)$ $L_2$ regularization. The error bar is based on the standard deviation over the five-fold cross validation. $(c)$ Streamwise velocity fields. The values underneath the contours represent the $L_2$ error norm.
}}
\label{fig_ROB-reg}
\end{center}
\end{figure}

The influence on the sparsity parameter $\alpha$ is investigated for both the $L_1$ and $L_2$ regularizations in figure~\ref{fig_ROB-reg}.
We have performed a five-fold cross validation for each $\alpha$~\citep{Bruntonkutz2019}, as presented as the error bar in figures~\ref{fig_ROB-reg}$(a)$ and $(b)$.
As a method for interpolation of lacked inputs, we use {\color{black}the cubic interpolation.}
The mean value of the $L_2$ error norm within the present cross-validation models do not show a clear dependence on the value of $\alpha$.
In contrast, focusing on the standard deviation, we can see that there is a large variation in the error depending on the hyperparameter $\alpha$.
Especially, the standard deviation with $\alpha = 1\times 10^{-6}$ is quite larger than the others, which suggests that one of the models with $\alpha = 1\times 10^{-6}$ exhibits a great robustness compared to the others.
The reason for the large variance in the error value is likely because the optimization of neural network is carried out in the high-dimensional solution space regarding the updating of the immense number of weights.
The same trend can also be found with the $L_2$ regularization in figure~\ref{fig_ROB-reg}$(b)$.
In sum, a cross validation is mandatory to reach the robust model.

\begin{figure}[t]
\begin{center}
\includegraphics[width=0.98\textwidth]{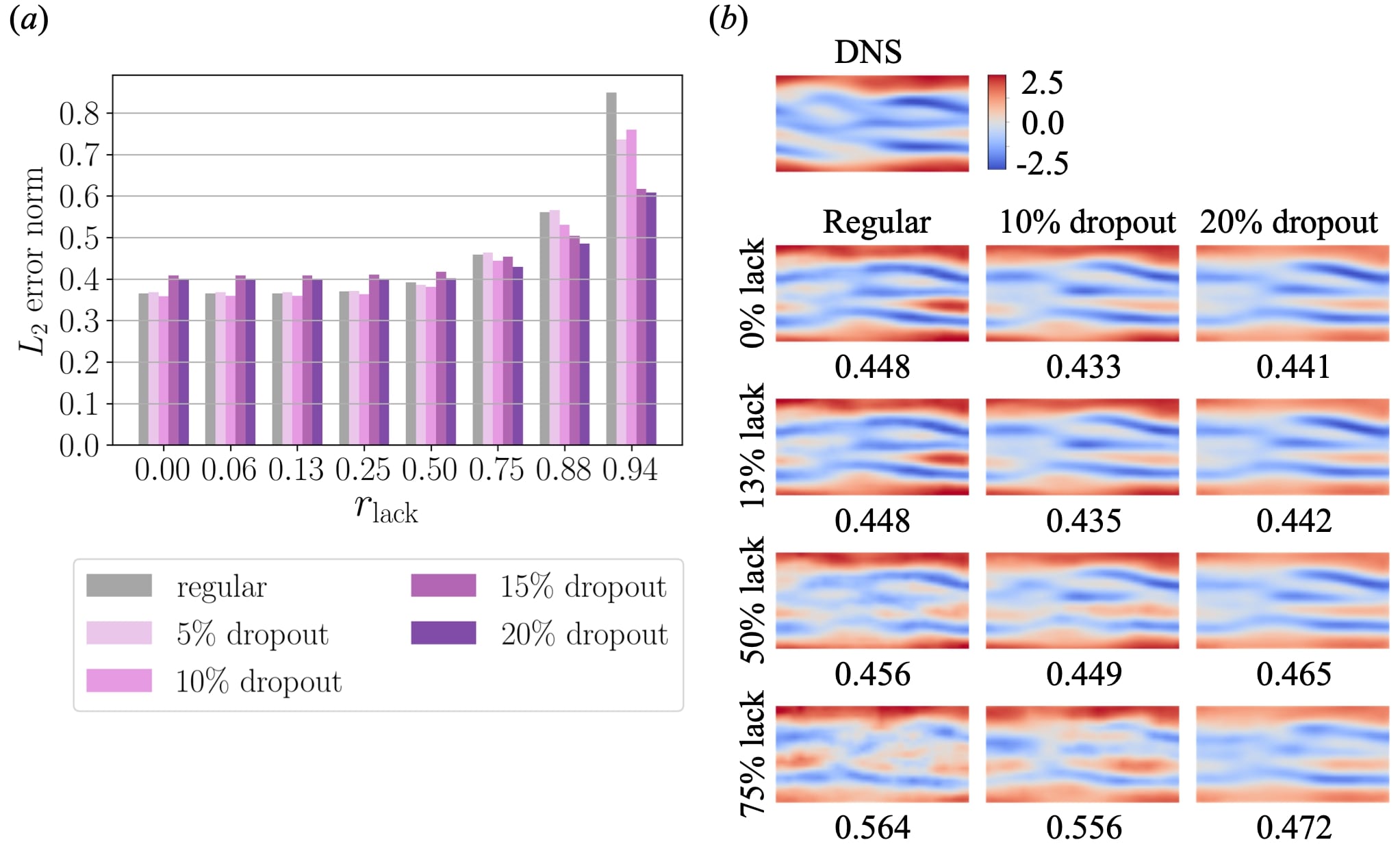}
\caption{{\color{black}
Effect of dropout for the robustness. The ``regular" is the model without dropout. $(a)$ Dependence of the $L_2$ error norm on the ratio of dropout and the sensor lack. $(b)$ Streamwise velocity fields. The values underneath the contours represent the $L_2$ error norm.}}
\label{fig_ROB-DO}
\end{center}
\end{figure}

The velocity fields for the regularization cases are summarized in figure~\ref{fig_ROB-reg}$(c)$.
Note that we here only visualize the best case of each $r_{\rm lack}$ since there is a high variation among the cross-validated models.
The regular CNN model without regularization is also shown for comparison.
With the regular model, the correct structures cannot be recovered with $75\%$ lack; however, this issue can be mitigated capitalizing on the regularization, especially with the $L_2$ regularization for the present case.
The superiority of the regularized models can also be observed from the $L_2$ error norm.
Summarizing above, the regularization is effective in obtaining robustness, although care should be taken for the choice of the hyperparameter~$\alpha$.
Constructing several models with the appropriate parameter~$\alpha$ results in a large variation in robustness, and users should choose the most robust model among them.

\begin{figure}[t]
\begin{center}
\includegraphics[width=0.98\textwidth]{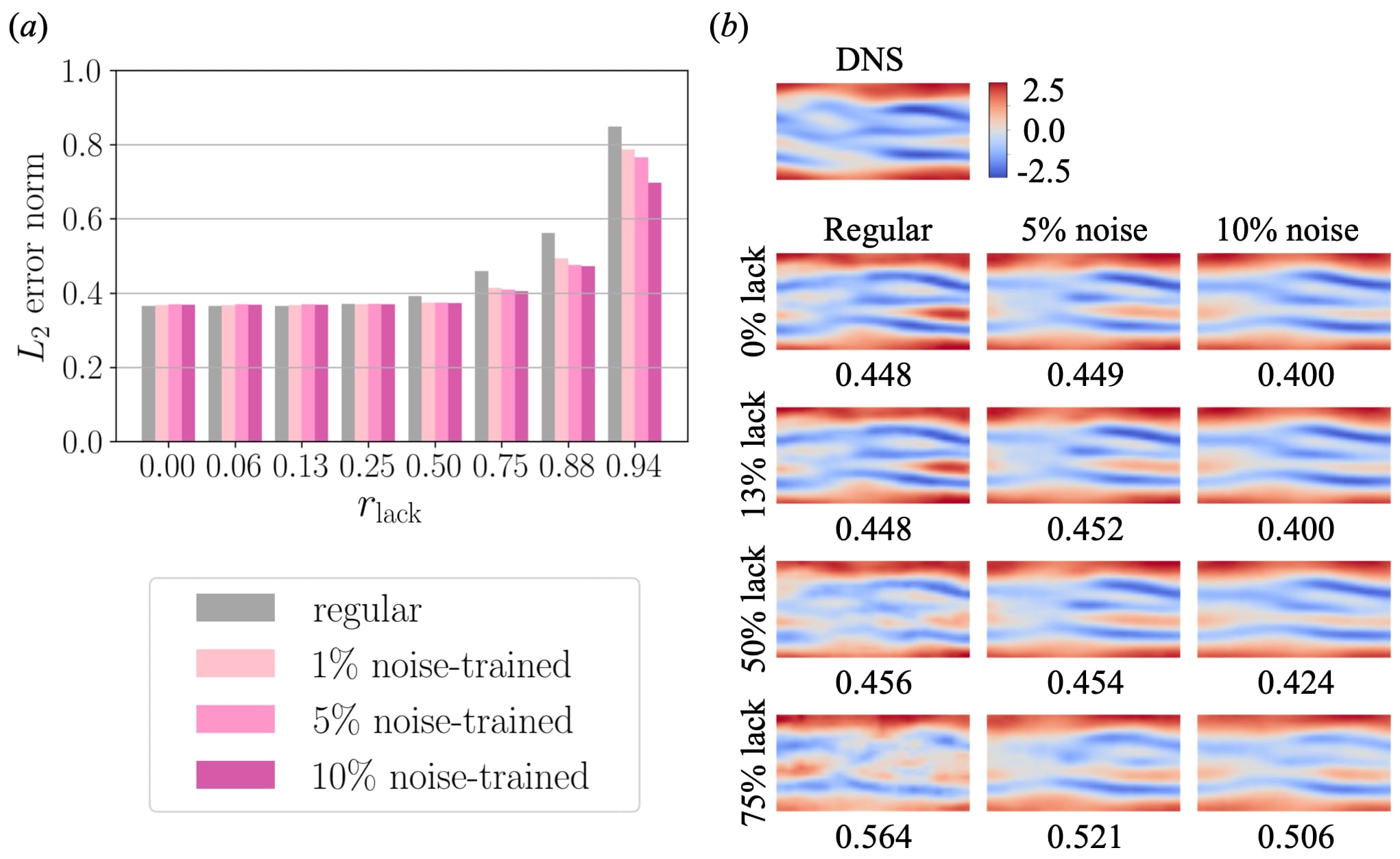}
\caption{{\color{black}
Effect of noise training for the robustness. The ``regular" is the model without noise training. $(a)$ Dependence of the $L_2$ error norm on the ratio of the noise for training data and the sensor lack. $(b)$ Streamwise velocity fields. The values underneath the contours represent the $L_2$ error norm.}}
\label{fig_ROB-noise}
\end{center}
\end{figure}

\subsubsection{Dropout}

Let us then consider dropout~\citep{SHKSS2014}, which is often used in machine learning to prevent overfitting.
This method can randomly deactivate a certain percentage of nodes during the training of a model, which essentially enables us to construct multiple models within a single model.
We examine the robustness for the lacked input by constructing several models whose dropout rates are different, as shown in figure~\ref{fig_ROB-DO}$(a)$.
The dropout rate here represents the deactivation rate of nodes in training.
The lack portion is interpolated using {\color{black}the cubic interpolation} analogous to the investigation of regularization.
The models with dropout outperform the regular model, especially with the large lack ratio.
Thus, a larger dropout ratio is indeed effective in obtaining robustness; however, we should note that too large dropout ratio deteriorates the model performance when the lack ratio is small since the connection inside the model is too sparse.
The velocity contours are also checked in figure~\ref{fig_ROB-DO}$(b)$.
Overall, we can observe the superiority of the dropout models over the regular one.
In particular, the flow field estimated by the model with $20\%$ dropout is in reasonable agreement with the DNS even with $75\%$ input.
Therefore, we can evaluate that dropout can make the CNN model robust for up to about $75\%$ lack in this example.

\subsubsection{Noise-addition training}

At last, we investigate the effect of noise-addition training.
Noise perturbation for training data is one of the data augmentation techniques to gain robustness of machine learning models because test data are generally treated as ``noise" against training data~\citep{SK2019,MFZF2020}.
We here consider a Gaussian noise whose magnitude is defined with signal-to-noise ratio (SNR), ${\rm SNR} = {\sigma^2_{\rm data}}/{\sigma^2_{\rm noise}}$, where $\sigma{^2}_{\rm data}$ and $\sigma{^2}_{\rm noise}$ are the variances of input data and noise, respectively.
The present study covers three different magnitudes of noise ${\rm 1/SNR}=\{0.01, 0.05, 0.10\}$.
The Gaussian noise is perturbed to 5000 snapshots out of 10000 training snapshots.
The machine learning model is constructed for each case and the training data sets (i.e., 5000 noisy snapshots plus 5000 clean snapshots) are also prepared for each case.

The error of each model for the lacked input is summarized in figure~\ref{fig_ROB-noise}$(a)$.
Comparing to the regular model, the noise-trained models are more robust for the lacked input, and the robustness is strengthened by adding a stronger noise though there is an upper limit.
The performance of the models is also verified using the streamwise velocity contours in figure~\ref{fig_ROB-noise}$(b)$.
The flow fields estimated by the models trained with noise-perturbed data are in good agreement with the DNS data even with the lacked input.
Noteworthy here is that the $L_2$ error against the $50\%$ lack of the model trained with $10\%$ noise is smaller than that against the $0\%$ lack of the model trained without noise.
This suggests that the noise perturbation to training data significantly helps the CNN model to obtain the strong robustness.

\section{Concluding remarks}
\label{sec5}

The practicability of neural network-based state estimation from limited sensor measurements in fluid flow was investigated.
We constructed estimation models utilizing convolutional neural network and applied to three types of unsteady laminar and turbulent flows which cover a wide range of spatial length scales associated with complex fluid flow phenomena.
The models were able to estimate a target of two-dimensional plane from input measurements.
From the viewpoint of a practicability, we also investigated physical quantities required for the input in the problems of turbulent channel flow and transitional boundary layer.
For both cases, the wall shear stress significantly contributed to the estimation performance.
The robustness of the models for the lacked input was further investigated towards the state estimation from much fewer available sensors.
We found that reasonable estimations can be achieved from the lacked input measurements by {\color{black}utilizing cubic interpolation.}
Moreover, the possibility for the utilization of several approaches for models to gain more robustness against a lack of sensors was demonstrated.

We can consider several {\it a posteriori} applications of the present robust fluid flow estimator based on neural network.
For example, it is expected that a machine learning-based estimator can help to control a flow by sensing and guessing a whole flow state~\citep{BN2015}.
In fact, the seminal work by~\citet{LKBG1997} used a shallow multi-layer perceptron to aid the opposition control~\citep{CMK1994} of the channel flow.
In addition to this study, several reports have demonstrated the applicability of the aforementioned combination based on the concept that estimates a velocity field on the detection plane from the wall measurements using a machine-learning model~\citep{han2020active,park2020machine,li2021blowing}.
However, it is also true that there are several remaining issues including the applicability of a model trained with uncontrolled cases to controlled flows in an online manner~\citep{park2020machine} and the limitation of the sensor availability in terms of both the number and the quality.
We believe that the present investigation can directly address these issues from the perspective on the robustness against noise and lack of the sensors.
As for the future study, the combination with the optimal sensor placements based on data-driven approaches~\citep{MBKB2018,saito2020data,NYNSN2020,morita2021applying} and digital twins~\citep{rasheed2020digital} can also be considered.

\paragraph{Acknowledgments}

We are grateful to Mr. Kai Fukami (UCLA) for fruitful discussion.
This work was supported by JSPS KAKENHI Grant Numbers 18H03758 and 21H05007.

\paragraph{Data availability}

The data that support the findings of this study are available from the corresponding author upon reasonable request.

\paragraph{Declaration of interest}

The authors report no conflict of interest.

\paragraph{CRediT Author contributions}

{\bf Taichi Nakamura:} Conceptualization, Methodology, Software, Validation, Formal analysis, Investigation, Data curation, Writing- Original draft preparation, Visualization.
{\bf Koji Fukagata:}
Conceptualization, Formal analysis, Investigation, Resources, Writing - Review \& Editing, Supervision, Project administration, Funding acquisition


\end{document}